\documentclass{IEEEtran}
\usepackage{cite}
\usepackage{amsmath,amssymb,amsfonts}
\usepackage{graphicx}
\usepackage{textcomp,nicefrac}
\usepackage{lineno}
\usepackage{wrapfig}
\usepackage{stfloats}

\def\BibTeX{{\rm B\kern-.05em{\sc i\kern-.025em b}\kern-.08em
		T\kern-.1667em\lower.7ex\hbox{E}\kern-.125emX}}
\begin{document}
	\title{Performance of the ATLAS Hadronic Tile Calorimeter Demonstrator system for the Phase-II upgrade facing the High-Luminosity LHC era.}
	
	\author{Eduardo Valdes Santurio	on behalf of the Tile Calorimeter System.
		\thanks{Eduardo Valdes Santurio, Physics Department (Fysikum), Stockholm University, Sweden. Email: eduardo.valdes@fysik.su.se, eduardo.valdes@cern.ch\\
			\textbf{This work was supported by Stockholm University and CERN.\\
			Copyright 2020, CERN, for the benefit of the ATLAS Collaboration. \\CC-BY-4.0 license}.
	}}

	
	\maketitle
	
	
	\begin{abstract}
		The High Luminosity Large Hadron Collider (HL-LHC) will have a peak luminosity of $5\times10^{34} $cm$^{-2} $\,s$^{-1}$, five times higher than the design luminosity of the LHC. The hadronic ATLAS Tile Calorimeter (TileCal)\cite{bib_atlas}\cite{bib_tilecal} is a sampling calorimeter with steel as absorber and plastic scintillators as active medium. The light produced in the scintillating tiles is guided to photomultiplier tubes (PMTs), where analogue signals are produced to be shaped and conditioned before being digitized every 25\,ns. TileCal Phase-II Upgrade for the HL-LHC will allow the system can cope with the increased radiation levels and out of time pileup. The upgraded system will digitize and send all the calorimeter sampled signals to the off-detector systems, where the events will be reconstructed and shipped to the first level of trigger, all at 40\,MHz rate. Consequently, development of more complex trigger algorithms will be possible with the more precise calorimeter signals provided to the trigger system. The new hardware comprises state of the art electronics with a redundant design and radiation hard electronics to avoid single points of failure, in addition to multi-Gbps optical links for the high volume of data transmission and Field Programmable Gate Arrays (FPGAs) to drive the logic functions of the off- and on-detector electronics. A hybrid demonstrator prototype module containing the new calorimeter module electronics, but still compatible with the present system was assembled and inserted in ATLAS during June 2019, so that the Phase-II system can be tested in real ATLAS conditions.
		
	\end{abstract}
	
	
	\section{Introduction}
\begin{figure}[!h]
	\centering
	\includegraphics[width=1.0\linewidth]{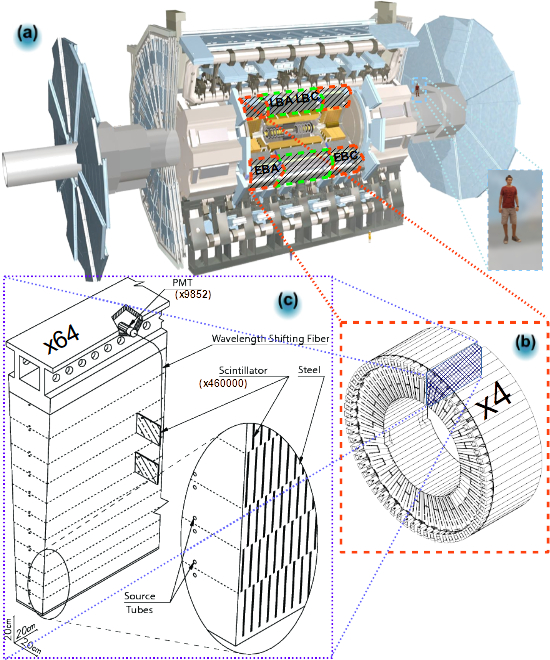}
	\caption{(a) The ATLAS detector. (b) A TileCal barrel. (c) Depiction of a TileCal wedge-shaped module.}
	\label{fig:atlas_tilecal}
\end{figure}

	The upgraded version of the Large Hadron Collider (LHC), HL-LHC, will have a peak luminosity of $5\times10^{34}$ cm$^{-2}$ s$^{-1}$,1 , which represents a five-fold increase in the designed luminosity of the LHC. The ATLAS experiment is a high energy physics general-purpose scientific apparatus instrumented with 34 multiple layers of detectors, that records the products of the hadron collisions. ATLAS TileCal (Figure \ref{fig:atlas_tilecal}) is a hadronic sampling calorimeter with steel as absorber and plastic scintillators as active medium. TileCal is longitudinally divided into three cylindrical barrels. Figures \ref{fig:atlas_tilecal}a and Figures \ref{fig:atlas_tilecal}b show the partitioning scheme of the barrels for TileCal, where the exterior barrels correspond to one partition each: the Extended Barrel A (EBA) and the Extended Barrel C (EBC); while the center barrel is divided into two partitions: Long Barrel A (LBA) and Long Barrel C (LBC). Each partition comprises $64$ wedge-shaped modules (Figure \ref{fig:atlas_tilecal}c). The light produced by charged particles in plastic scintillating tiles is collected on each side of a pseudo-projective cell by wavelength shifting fibers and read out by a pair of photomultiplier tubes (PMTs). Each PMT produces analogue signals that are then shaped, conditioned and amplified into two gains, before being digitized every $25$\,ns. 

	The Phase-II Upgrade of TileCal for the HL-LHC presents a system that can cope with the HL-LHC expected increase of radiation levels. The upgraded electronics continuously read out digital data of all TileCal channels with better energy resolution and less sensitivity to out-of-time pileup. The state of the art electronics used for the new design provides better timing stability and lower electronic noise \cite{bib_tilecal_phase_ii_tdr:2018}. Pipeline memories will sit on the off-detector systems, where the events will be reconstructed and transmitted to the first level of trigger, all at 40 MHz rate. This scheme will allow further development of more complex trigger algorithms with the more precise calorimeter signals provided to the trigger system. The upgraded hardware features state of the art electronics crafted into a design with redundancy and radiation tolerant strategies to avoid single points of failure. Multi-Gbps optical links are used for the high volume of data transmission and Field Programmable Gate Arrays (FPGAs) will manage the high volume of digital processing of the off- and on-detector electronics. A hybrid module with the new calorimeter electronics and compatible with the present system, called the Demonstrator, was assembled to follow-up the R\&D work for TileCal Phace-II Upgrade. The Demonstrator was extensively tested during five testbeam campaigns \cite{bib_testbeam}, so that it could be inserted in ATLAS for further studies of the Phase-II upgrade electronics system in real ATLAS conditions.

\section{The Demonstrator read-out system}
\begin{figure}[!ht]
	\centering
	\includegraphics[width=1.05\linewidth]{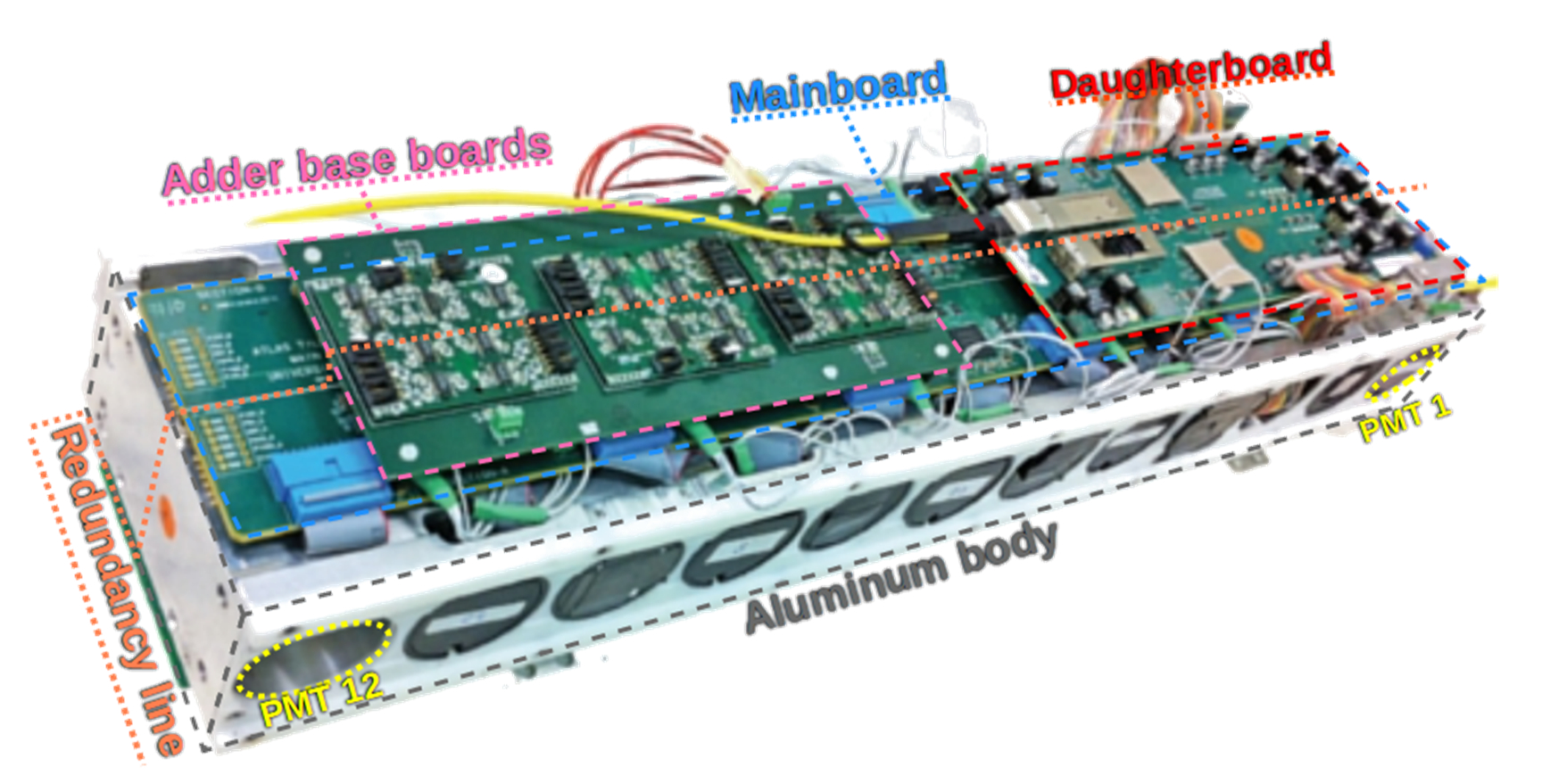}
	\caption{TileCal hybrid Demonstrator Minidrawer.}
	\label{fig:md}
\end{figure}

\begin{figure*}[b!]
	\centering
	\includegraphics[width=0.9\linewidth, height=5cm]{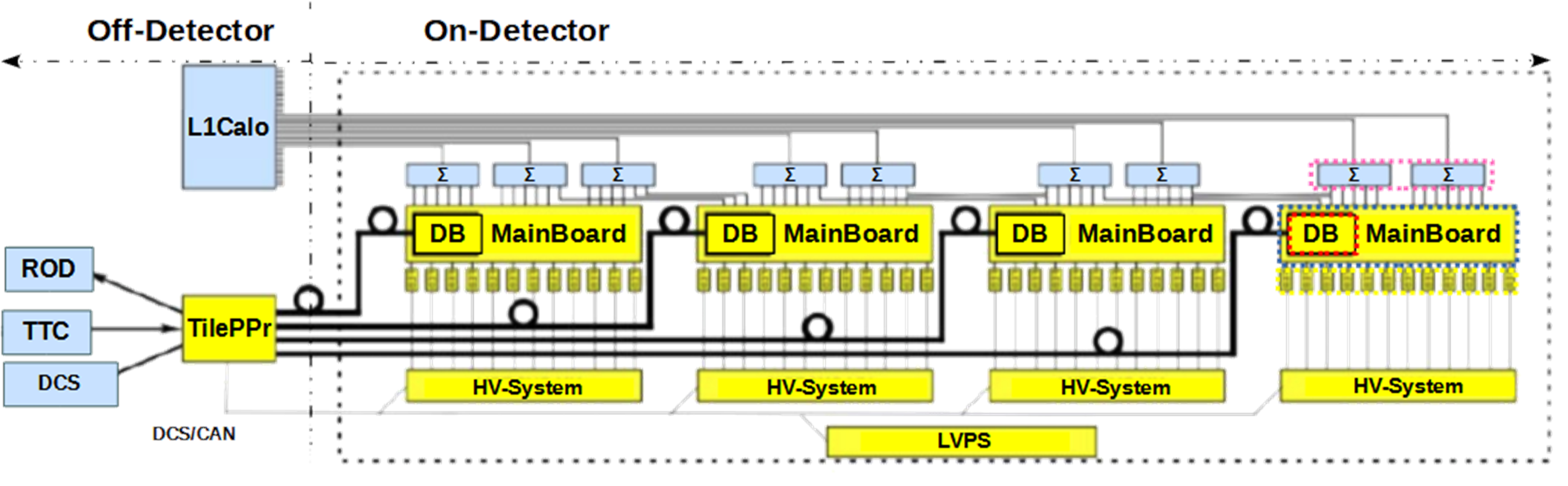}
	\caption{Block diagram of the Demonstrator read-out system (yellow) and interface with TileCal legacy electronics (blue).}
	\label{fig:demonstrator_diagram}
\end{figure*}

The Demonstrator on-detector read-out electronics consists of a so called Superdrawer (SD) that partitions a legacy TileCal Drawer into four Minidrawers (MD, Figure \ref{fig:md}), each servicing up to 12 PMT channels. The SD will continuously digitize two gains of up to 48 TileCal PMTs and send the digitized sampled data to the off-detector systems at 40 MHz (Figure \ref{fig:demonstrator_diagram}) \cite{bib_tilecal_demonstrator:2017}. 

Each PMT analogue signal is conditioned, shaped and fed into amplifiers with low- and high-gain of ratio 1:32 by a 3-in-1 card, before they are digitized by a Mainboard. The 3-in-1 card is an early version of the Front End board for the New Infrastructure with Calibration and signal Shaping (FENICS) card, made to be compatible with the legacy TileCal trigger system. A Daughterboard (DB) receives and propagates LHC synchronized timing, control signals and configuration commands to the front-end, while transmitting continuous read-out from the MB channels to the off-detector systems. Compatibility with the legacy trigger system of TileCal is achieved by an Adder-based-board ($\Sigma$) that groups the PMT analogue in cell pseudo-projective towers and send analogue sums to the legacy L1Calo trigger system.

On the Off-Detector electronics, a Tile Preprocessor (TilePPr) continuously receives and stores the PMT data of the entire SD in pipelines until the reception of a trigger decision event, where corresponding data is transferred to the legacy ROD (Read Out Driver). The TilePPr serves as a hub between the legacy system and the new electronics by propagating LHC synchronous clocks and configuration commands to the each of the MDs, receiving commands and trigger signals from the Trigger, Timing and Control (TTC), interfacing with the Detector Control Systems (DCS), and sending the triggered data to the legacy ROD.


\section{The Demonstrator intagration, status and test results}

\begin{figure}[h!]
	\centering
	\includegraphics[width=1.00\linewidth]{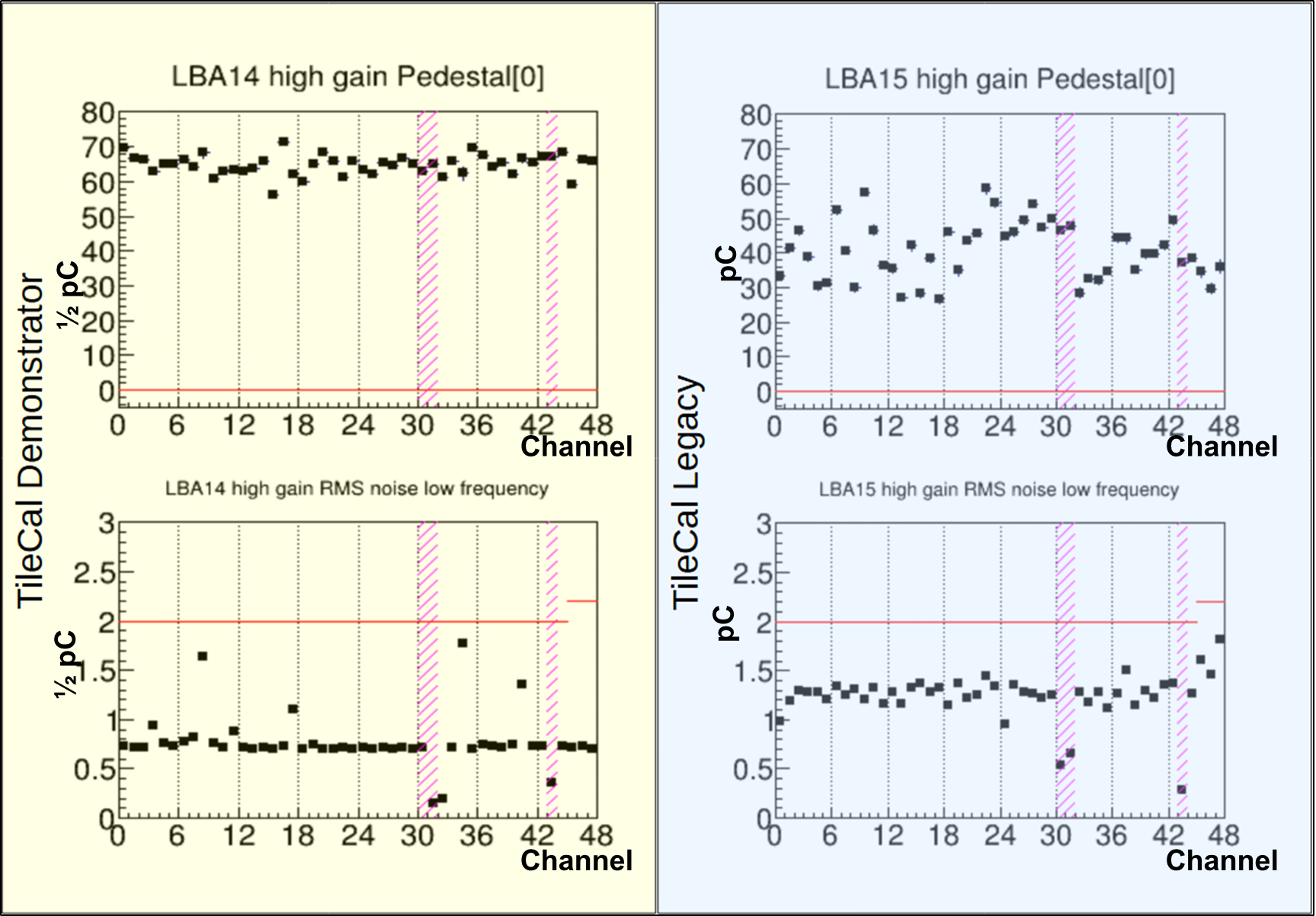}
	\caption{High gain pedestals and noise from the Optimal Filter. To the left (yellow): Demonstrator Pedestal and RMS. To the right (blue): TileCal legacy drawer Pedestal and RMS.}
	\label{fig:pedestal}
\end{figure}

\begin{figure}[h!]
	\centering
	\includegraphics[width=1.00\linewidth]{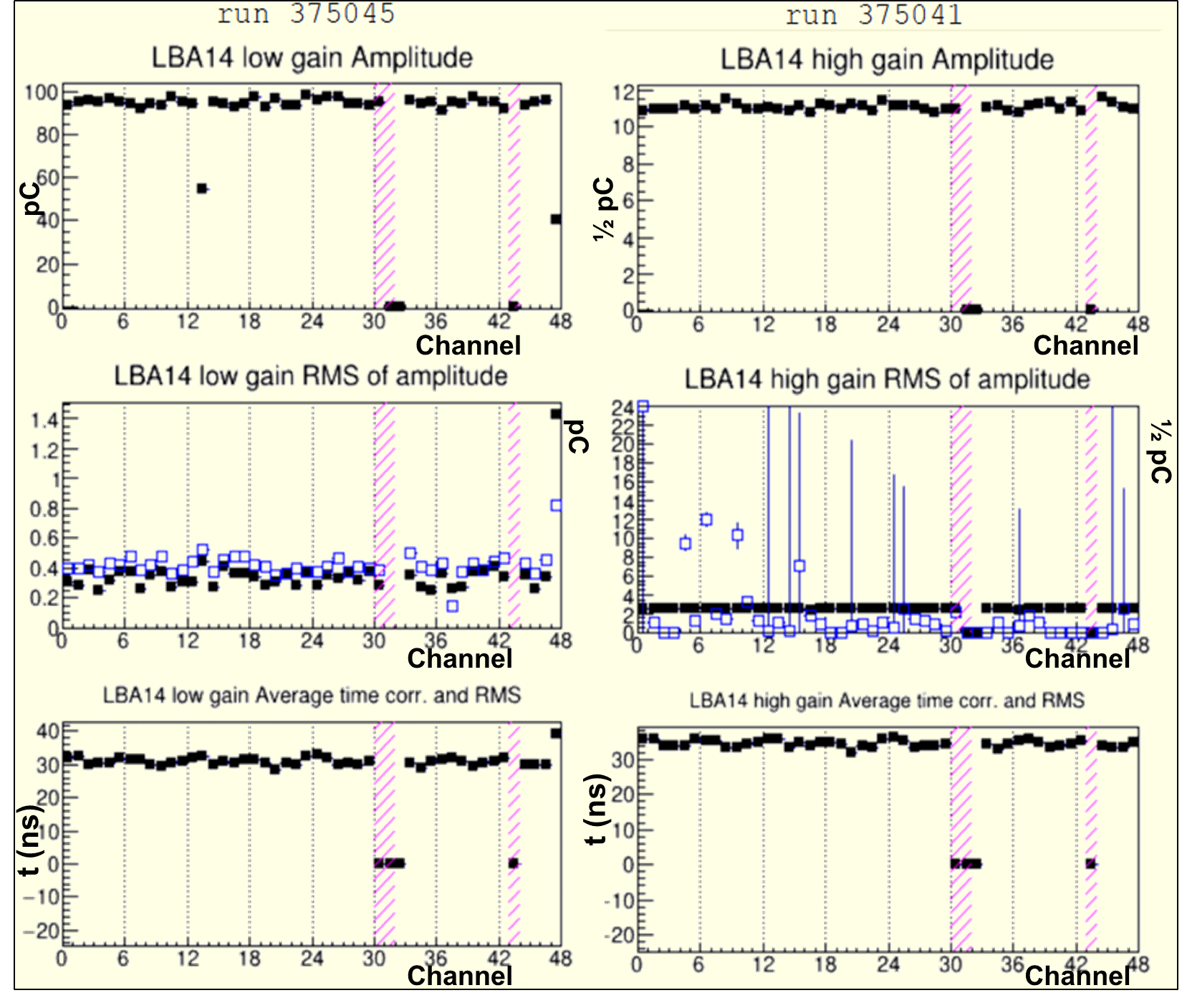}
	\caption{Charge Injection System runs from the Optimal Filter. To the left: Demonstrator Low Gain CIS run amplitudes, RMS and average time correlation. To the right (blue): Demonstrator Low Gain CIS run amplitudes, RMS and average time correlation.}
	\label{fig:cis}
\end{figure}

\begin{figure}[h!]
	\centering
	\includegraphics[width=1.025\linewidth]{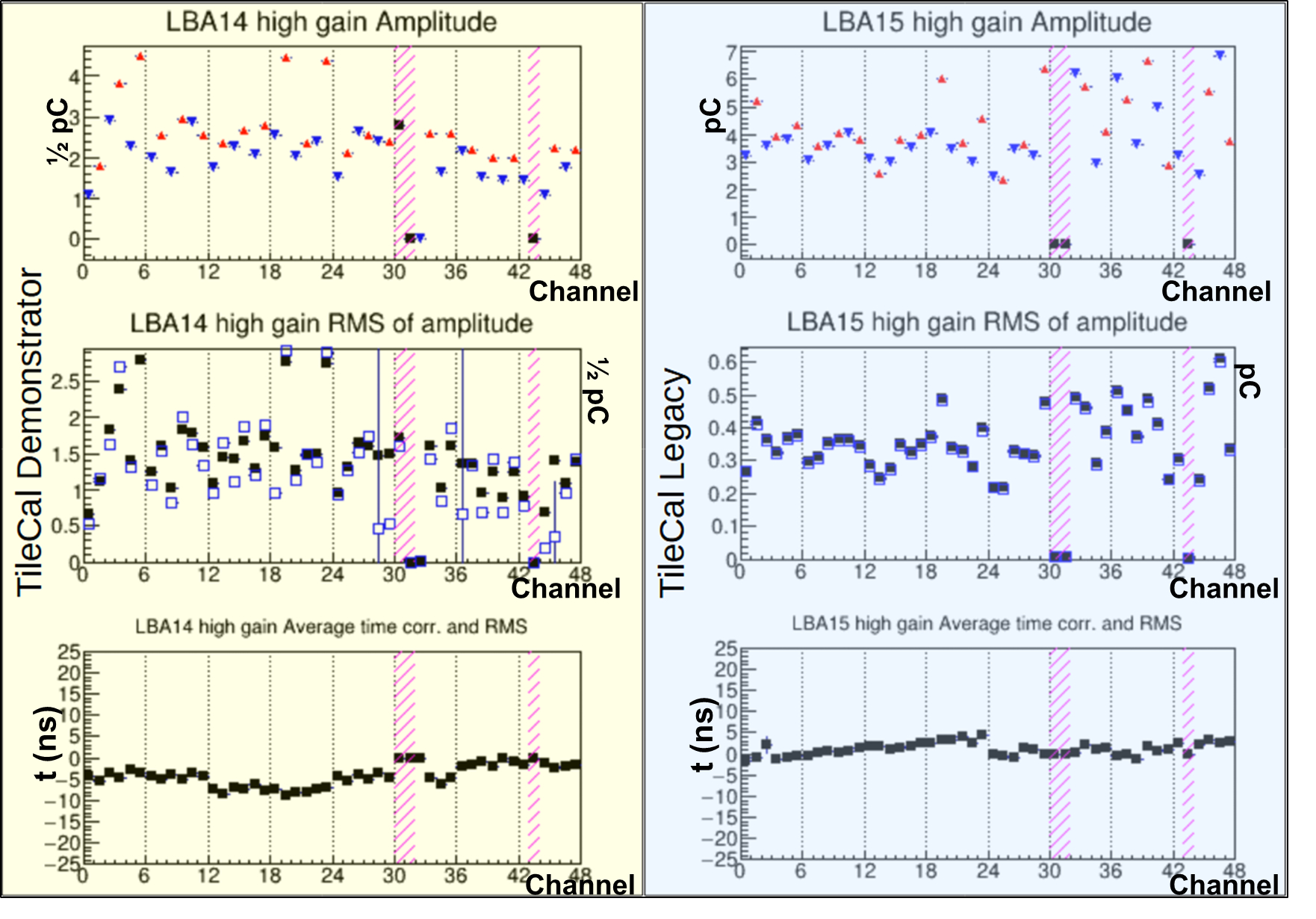}
	\caption{High gain Laser runs from the Optimal Filter. To the left (yellow): Demonstrator Laser run amplitudes, RMS and average time correlation. To the right (blue): TileCal legacy drawer Laser run amplitudes, RMS and average time correlation.}
	\label{fig:laser}
\end{figure}

A Demonstrator with Phase-II electronics was successfully installed in ATLAS and interfaced with the TileCal legacy system. The system exhibited very stable links between the on-detector electronics sitting in the ATLAS cavern and the off-detector electronics sitting at the counting room (USA15). The Demonstrator read-out was fully integrated with the TDAQ software used for the legacy system. Notice in Figures \ref{fig:pedestal}, \ref{fig:laser} and \ref{fig:cis} that the current values reported by the TDAQ software for the read-out of the high-gain data of the Demonstrator corresponds to half dimension of the legacy system read-out value, as pointed out in the y-axis of the correspondent charts. This is caused by the difference between gain ratios of 1:32 and 1:64 for the Demonstrator and the legacy system respectively not being implemented on the legacy TDAQ software. The DCS was interfaced with the Demonstrator module to power and monitor the LVPS and HV system. A set of sensors were interfaced to the MDs to monitor the temperature and humidity on critical positions such as FPGAs, GBTx ASICs, DC-DC converters and PMTs inside of the TileCal girder. These measurements were within the ranges specified by the respective manufacturers during the Demonstrator normal operations.

The Demonstrator read-out has a 17-bit dynamic range achieved by two-gain 12-bit fast ADC read-out with ratio of 1:32, in contrast to the 16-bit dynamic range provided by the two-gain ADC read-out with ratio of 1:64 of the legacy drawer. In the demonstrator MBs, high precision Digital to Analogue Converters (DACs) can be used to feed a desired voltage level, or pedestal, to each of the ADC inputs. The pedestal tests consist in configuring the DACs to determined values, that will be seen as a constant offset in the read-out when no pulses are present or at the sides of a triggered pulse. The pedestal tests ran in both systems showed lower noise in the demonstrator, which was expected due to the larger dynamic range featured of the Demonstrator. Figure \ref{fig:pedestal} shows a side-to-side comparison between pedestals and noise from the Optimal Filtering algorithm (OF) for all the channels of the Demonstrator and a TileCal legacy drawer.

The Demonstrator calibration and test suite includes a Charge Injection System (CIS). Electronic pulses with configurable amplitude can be synchronously injected into the shaper-amplifier stage of the FENICs cards to test the timing and behaviour of the read-out chain of each channel from the amplifiers and up. Timing calibration with CIS runs were successfully performed for low and high gains on the Demonstrator module (Figure \ref{fig:cis}).

The Laser system distributes synchronous short and controlled light pulses over optical fibres to the light mixers of the PMT blocks. This allows measurements of individual timing and behaviour of the read-out chain from each PMT and up. 
The Laser run tests (Figure \ref{fig:laser}) showed comparable performance between the Demonstrator in amplitudes, amplitude RMS and average time correlations. The average time correlations are calculated through computation of the phase difference between the fitted pulse and the ideal pulse with reference to the forth sample out of the 7 samples used by the OF algorithm in TileCal.

The dark current measurement where performed by measuring the difference between pedestals with the HV switched ON and the pedestals with the HV switched OFF, namely $\Delta$(Ped). The measurements were run under conditions where light leakage and factors such as radiation source, activation, or beam background are all negligible. Therefore, $\Delta$(Ped) can be attributed mostly to the dark currents. The dark current measurements showed no significant difference in gain measurements (Average for all channels: $\Delta$(Ped) = 47\,ADC =  29\,pA) (Figure 7 - measurement for channel 0). During the tests it was noticed that the integrator pedestals are in general lower when the DAC is disconnected compared to when it is connected and set to zero.

\begin{figure}[h!]
	\centering
	\includegraphics[width=1.00\linewidth]{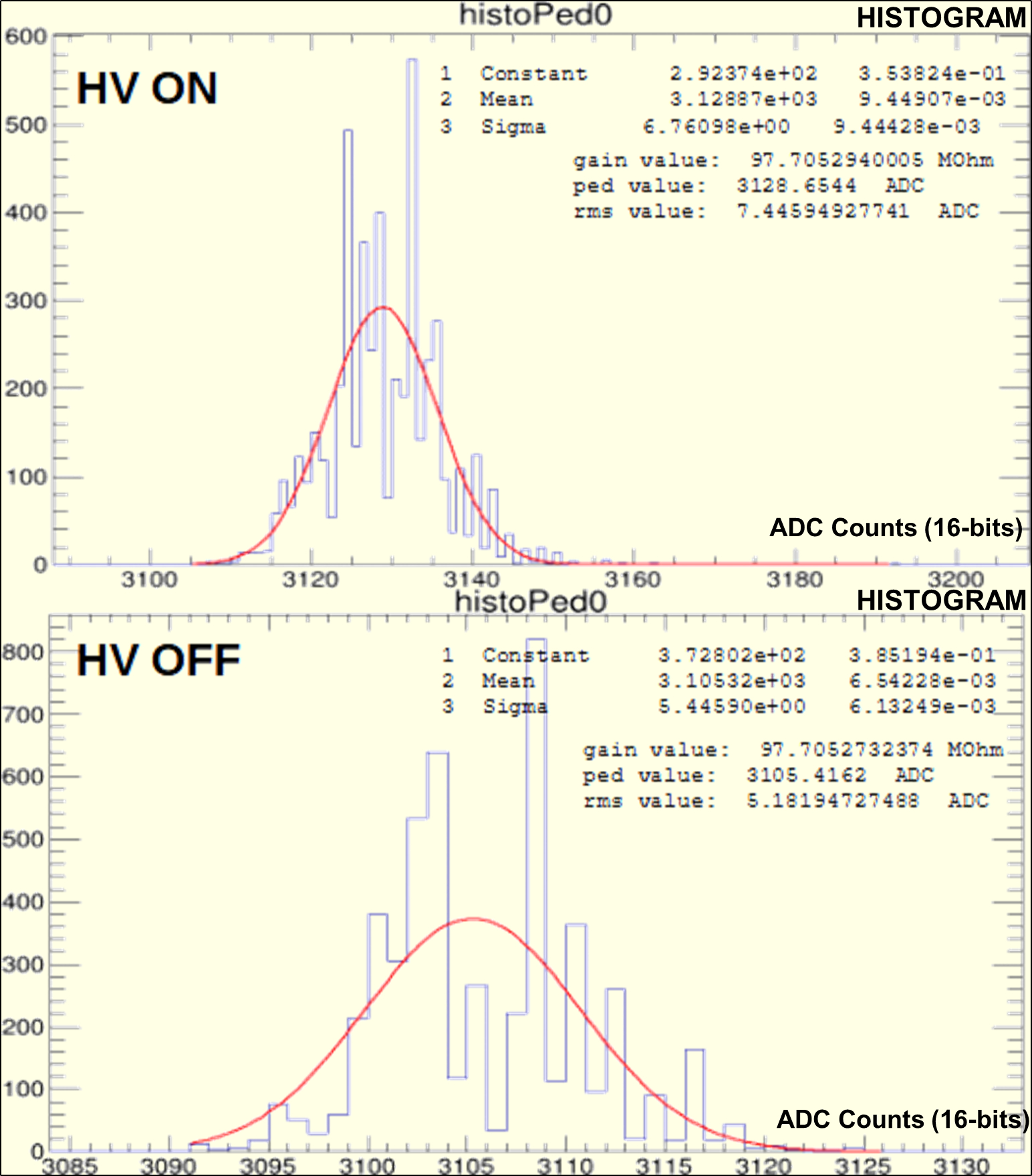}
	\caption{High gain Laser runs from the Optimal Filter. To the left (yellow): Demonstrator Laser run amplitudes, RMS and average time correlation. To the right (blue): TileCal legacy drawer Laser run amplitudes, RMS and average time correlation.}
	\label{fig:dark_currents_v1}
\end{figure}

\section{Conclusions}
The demonstrator is fully integrated with the TileCal Legacy TDAQ, and DQ plots are available for it each time a calibration is taken. However, more DQ tests and tuning need to take place (e.g. using channels with unattached PMTs, called PMT holes). The Demonstrator timing was adjusted by CIS and laser, and the integrator pedestals are in general lower when the DAC is disconnected compared to when it is connected and set to zero. No significant difference could be seen when the HV is ON vs OFF. The Demonstrator data acquisition runs have taken place without data losses. Future plans include more integration and further studies with new optimized TilePPr firmware versions. The Demonstrator activities will prove reliability and provide valuable experience for the future installation of the HL-HLC upgraded electronics. The fact that the Demonstrator is integrated in the ATLAS environment has provided valuable experiences and useful information that will help the future installation of the HL-LHC upgraded electronics.
	
	\section*{References}
	
	\def\refname{\vadjust{\vspace*{-1em}}} 


\begin{thebibliography}{9} 
		
	\bibitem{bib_atlas}
	ATLAS Collaboration, 2008 JINST 3 S08003.

	\bibitem{bib_tilecal}	
	ATLAS Collaboration, ATLAS Tile Calorimeter : Technical Design Report, CERN-LHCC-96-042, ATLAS-TDR-3, 1996.
	
	\bibitem{bib_tilecal_phase_ii_tdr:2018}
	ATLAS Collaboration. Technical Design Report for the Phase-II Upgrade of the ATLAS Tile Calorimeter. CERN-LHCC-2017-019, ATLAS-TDR-028, 2018.

	\bibitem{bib_tilecal_demonstrator:2017}
	E. Valdes Santurio, ATLAS Collaboration, Upgrade of Tile Calorimeter of the ATLAS Detector for the High Luminosity LHC, Journal of Physics: Conference Series, DOI: 10.1088/1742-6596/928/1/012024.
	
	\bibitem{bib_testbeam}
	E. Valdes Santurio, ATLAS Collaboration, Beam Tests on the ATLAS Tile Calorimeter Demonstrator Module, urn:nbn:se:su:diva-175281, 10.1016/j.nima.2018.10.066.
	

	
%
%
%
%
%
	
	
	
	
	
\end{thebibliography}
\end{document}